# Teacher's Evaluation – a Component of Quality Assessment System

**Assoc.Prof. Ph.D. Tiberiu Marius KARNYANSZKY**
**Assoc.Prof. Ph.D. Laurenţiu Dan LACRĂMĂ**
**Assoc.Prof. Ph.D. Lucian LUCA**
**Univ.Assist. Ioana IACOB**
"Tibiscus" University of Timişoara, România

ABSTRACT. One of the most important activities to increase the importance and the responsibility of the higher education is the quality management, assessment and evaluation. Starting from 2006, a national mechanism was created in Romania and all the educational institutions have to apply a concrete algorithm to ensure the internal evaluation, the external evaluation and, the most important, to increase the quality of the educational process.
This paper presents the implementation of the quality assessment in "Tibiscus" University of Timişoara, particularly at the Faculty of Computers and Applied Computer Science.

## 1  The Principles of the Evaluation

activity of periodical evaluation of the teaching staff from our faculty is based on the stipulations of the specifications of The Romanian Agency for Quality Assurance in Higher Education (Agenţia Română de Asigurare a Calităţii în Învăţământul Superior, ARACIS) namely[1]:
*„The quality of the teaching and researching staff*
The universities must dispose of teaching staff which, as number and functional base, must be correctly allocated to the total number of students, depending on the study domain, and regarding the qualifications it must

---

[1] [Ara06] p. 37

depend on the specific of the study program and the proposed quality objectives."

According to these principles, the evaluation procedure elaborated by the Commission of Quality Assessment and Assurance in the "Tibiscus" University (Comisia de Evaluare şi Asigurare a Calităţii în Universitatea „Tibiscus", CEACUT) and applied by the faculty Quality Commission (Comisia Calităţii, CC), count on the following algorithm: every teacher from the Computer Science Chair applies a quality management evaluation procedure:

- The student's evaluation;
- The self evaluation;
- The collegial evaluation;
- The head of chair's evaluation.

## 2    The Student's Evaluation

The regulations regarding the periodical evaluation of the staff are presented in the ARACIS standards as[2]:

„*The students' evaluation of the teaching staff*

Minimum standard: A questionnaire for the students' evaluation of the teaching staff is available, it is approved by the university senate, it's applied on demand at the end of each semester and the application's results are confidential and accessible only to the dean, the rector and the evaluated teacher.

Optimal standard: The evaluation is compulsory. The results of the student's evaluation are separately discussed, statistically processed apart on chairs, faculties and university, to obtain a transparent policy on the educational quality."

This evaluation is based on a questionnaire developed by the Faculty of Psychology and contains 58 questions regarding:

- The scientific competence (12 items);
- The psycho pedagogical competence (20 items);
- The psychosocial competence (13 items);
- The managerial competence (13 items).

Students must complete the questionnaire and use the following answers:

---

[2] [Ara06] p. 38

- Teacher always harmonizes to the question (5 points)
- Teacher almost harmonizes to the question (4 points)
- Teacher neither harmonizes nor discords (3 points)
- Teacher almost discords to the question (2 points)
- Teacher always discords to the question (1 point)

Only the complete questionnaires are memorized into the answers database. Answers are concentrated on the four competencies and a media of the answers is calculated. If necessary, the resulting points are translated into marks expressed by words (4.01-5.00 means „very good", 3.01-4.00 means „good", 2.01-3.00 means „medium", 1.01-2.00 means „poor", 0-1.00 means „very poor").

For the 2006/2007 academic year, the results of the processed questionnaires are:

- The scientific competence – between 3,8 and 4,2;
- The psycho pedagogical competence – between 4,11 and 4,6;
- The psychosocial competence – between 3,98 and 4,50;
- The managerial competence – between 4,12 and 4,40.

Starting with 2007, the teachers evaluation is based on a computer application as presented in [TKS08].

## 3      The Self Evaluation

Use the same questionnaire as used for the students evaluation. Each evaluated teacher must complete the questionnaire and major differences between both evaluations are debated with the chief of staff.

For the 2006/2007 academic year, the results of the self evaluations are:
- The scientific competence – between 3,85 and 4,26;
- The psycho pedagogical competence – between 4,2 and 4,68;
- The psychosocial competence – between 4,05 and 4,6;
- The managerial competence – between 4,14 and 4,48.

## 4      The Collegial Evaluation

The regulations regarding the periodical evaluation of the staff are presented in the ARACIS standards as[3]:

---

[3] [Ara06] pp. 37-38

*"Collegial evaluation*
Minimum standard: The fellow-like evaluation is periodical and based on general criteria and also on collegial preferences.
Optimal standard: The fellow-like is compulsory and periodical. Each chair and department has a commission for the evaluation of the didactic and the research evaluation of each teacher/researcher; this commission elaborates an annual report on the staff's didactical and researcher quality."

The aim of the collegial evaluation is to offer to the teacher information to increase his teaching methods.

The evaluation criteria are:
- the teaching activity content
- the lesson's organization and presentation
- the lesson's clarity
- the connection to the students

The participation to the evaluation is compulsory for the evaluated teacher and voluntary for the evaluators. The observation during the courses (seminary/laboratory) was managed by a fellow, appointed by the chief of staff/dean, according to the evaluator and the evaluated person's opinion.

According to the psychological educational principles, the evaluation has three steps: pre-observation, observation and post-observation. These include a discussion about the progress of the lesson, the lesson's moments, the lesson's content, and the conclusions.

## 5     The Chief of Staff's Evaluation

The chief of staff's evaluation has, in our opinion and our algorithm, two steps: the evaluation based on the teacher's activity and a final evaluation based on all previous evaluations.

The chief of staff evaluates the teacher's performances regarding to:
- Didactical materials writhed (university courses or lectures, on printed/electronically recorded/Internet accessible)
- Scientific research (papers, articles, presentations at national and international conferences, published in national or international evaluated journals or proceedings, peer-to-peer or blind evaluations)
- Work with the students (papers, students study groups, collaboration into projects)
- National recognition (awards, decorations, fellowships to professional associations)

- International recognition (awards, decorations, inclusions into scientists databases, fellowships to professional associations)
- Work in the academic community (participation to local projects, eligible functions, committees)
- Participation to the institutional development (committees, extracurricular work)

The criteria weight depends on teachers' title, the scientific research and the work recognition increasing for higher titles:

| No | Evaluation criteria | Criteria weight | | | | |
|---|---|---|---|---|---|---|
| | | Prof. | Assoc. Prof. | Assist. Prof. | Univ. Assist. | Instructor |
| 1 | Elaboration of didactic materials | 10 % | 15 % | 10 % | 5 % | 5 % |
| 2 | Scientific research | 30 % | 25 % | 25 % | 25 % | 25 % |
| 3 | Activity with the students | 10 % | 10 % | 20 % | 25 % | 25 % |
| 4 | National Recognition | 10 % | 20 % | 15 % | 5 % | 5 % |
| 5 | International recognition | 15 % | 5 % | | | |
| 6 | Activity in the academic community | 10 % | 10 % | 15 % | 30 % | 30 % |
| 7 | Participation to institutional development | 15 % | 15 % | 15 % | 10 % | 10D% |

The chief of staff's final evaluation depends on evaluates teacher's performance regarding to the student's evaluation, self evaluation, fellow-like evaluation and chief of staff's evaluation like in the following example:

| No | Criteria | Qualifficatives | Marks |
|---|---|---|---|
| 1 | Fellow-like evaluation | Very Good | - |
| 2 | Chief of staff's evaluation | - | 4,50 |
| 3 | Student's evaluation | - | 4,11 |
| 4 | Self evaluation | - | 4,20 |
| | **FINAL RESULT** | **Very Good** | |

where the mathematical media of all evaluations determines the final qualificative.

**Conclusions**

This paper presents the teacher's evaluation system, applied to the "Tibiscus" University of Timişoara. This mechanism is based on the legal demands of the Romanian Agency for Quality Assurance in Higher Education. This evaluation is based on 4 steps: student's evaluation, self evaluation, collegial evaluation, head of staff's evaluation. Each evaluation offers a mark (Very Good, Good, Medium, Poor, Very Poor) and the media of this marks is used for the final qualificative.